%% file: colm2026_conference.tex
\documentclass{article} 
\usepackage[preprint]{colm2026_conference}

\input{math_commands.tex}

\usepackage{microtype}
\usepackage{hyperref}
\usepackage{url}
\usepackage{graphicx}
\usepackage{booktabs}
\usepackage{array}
\usepackage{multirow}
\usepackage{tikz}
\usepackage{pgfplots}
\usepackage{listings}
\usepackage{float}
\usepackage{colortbl}
\usepackage{wrapfig}
\usepackage{algorithm}
\usepackage{algpseudocode}
\usepackage{xcolor}
\usepackage{lineno}
\usepackage{comment}

\definecolor{darkblue}{rgb}{0, 0, 0.5}
\hypersetup{colorlinks=true, citecolor=darkblue, linkcolor=darkblue, urlcolor=darkblue}

\title{SynSQL: Synthesizing Relational Databases for Robust Evaluation of Text-to-SQL Systems}

\author{Mohammadamin Habibollah, Davood Rafiei\\
Department of Computing Science\\
University of Alberta\\
\texttt{\{mhabibol, drafiei\}@ualberta.ca}
}

\newcommand{\EXc}{\ensuremath{\mathrm{EX}_{c}}} 

\begin{document}

\ifcolmsubmission
\linenumbers
\fi

\maketitle

\begin{abstract}

   
   Evaluating text-to-SQL systems remains largely fragile: correctness is typically judged by executing predicted and gold SQL queries on a single static database, even though the same queries may behave differently under alternative database instances. This raises a broader language modeling question: \textbf{Can large language models synthesize semantically meaningful, schema-consistent relational data directly from a natural language question?} If so, such generation can serve as a controlled mechanism for stress-testing text-to-SQL systems beyond fixed benchmark databases.
   We introduce SynSQL, a framework that synthesizes test databases conditioned on question–schema alignment rather than gold SQL queries. SynSQL decomposes the task into three stages: (1) schema selection, (2) question-guided data synthesis, and (3) constraint-aware critique with iterative refinement, framing database construction as structured generation under semantic and relational constraints. 
   Across ten text-to-SQL models on Spider, BIRD, and Spider 2.0, SynSQL-generated databases reveal performance drops of 3–14\% compared to static evaluation, exposing errors masked by benchmark artifacts. We further analyze generation quality, constraint adherence, and failure modes, highlighting both the promise and limitations of LLMs in structured data synthesis. Our findings position synthetic database generation as a new lens for studying LLM reasoning, controllability, and robustness in structured environments.
   
   \end{abstract}
   
   \section{Introduction}
   
   
   Text-to-SQL benchmarks such as Spider~\citep{yu2018spider} and BIRD~\citep{li2023can} have driven rapid progress by pairing natural language (NL) questions with gold SQL queries over curated relational databases. Evaluation is typically conducted by executing predicted and gold queries against a single fixed database instance and comparing their outputs. While effective for standardization, this setup implicitly assumes that correctness is invariant to the underlying database content. 
   In practice, this assumption is fragile, and execution outcomes can depend heavily on the specific database instance~\citep{mitsopoulou2025analysis,renggli2025fundamental}. Issues such as referential integrity violations, unexpected \texttt{NULL} values, case mismatches between questions and database content, or noisy entries may cause incorrect queries to appear correct (false positives) or semantically valid queries to be penalized (false negatives). As a result, evaluation reflects not only language understanding and query reasoning, but also the particular ``world'' encoded by the benchmark database~\citep{zhong2020semantic}.
   
   This fragility suggests a broader perspective: evaluation can be viewed as a data generation problem. Instead of asking whether a predicted query matches a gold query on a single database, we should ask whether it remains correct across  semantically meaningful variations of the underlying data, an important but underexplored aspect of text-to-SQL evaluation. Prior approaches to generating alternative databases are gold query-centric. Systems such as AGENDA~\citep{deng2005testing}, XData~\citep{veanes2010qex}, and TestSuiteAccuracy~\citep{zhong2020semantic} construct counterexample databases by mutating gold SQL queries or by symbolically analyzing them to distinguish correct from incorrect variants. More recent SMT-based approaches (e.g., VeriEQL~\citep{he2024verieql} and SpotIt~\citep{klopfenstein2025spotit}) synthesize databases that distinguish predicted queries from reference queries. While powerful, these methods rely on access to gold SQL and are limited by bounded verification and query complexity. More fundamentally, they leave open a key question: \emph{can relational test data be generated directly from natural language and schema structure?}
   
   In this work, we investigate whether large language models (LLMs) can synthesize semantically grounded, schema-consistent relational databases conditioned only on a natural language question and a database schema. If feasible, such question-conditioned generation serves two purposes. First, it enables robustness evaluation of text-to-SQL systems without relying on gold query annotations and their mutations.
   Second, it probes an emerging capability of LLMs: structured world modeling under relational constraints. Generating a database from a question requires identifying relevant schema elements, populating tables with coherent and discriminative values, enforcing integrity constraints (e.g., foreign keys and uniqueness), and encoding semantic signals that expose query errors. This reframes database synthesis as a controlled structured generation task that jointly tests semantic grounding and constraint awareness.

   \begin{figure*}[t]
   \centering
   \includegraphics[width=\textwidth]{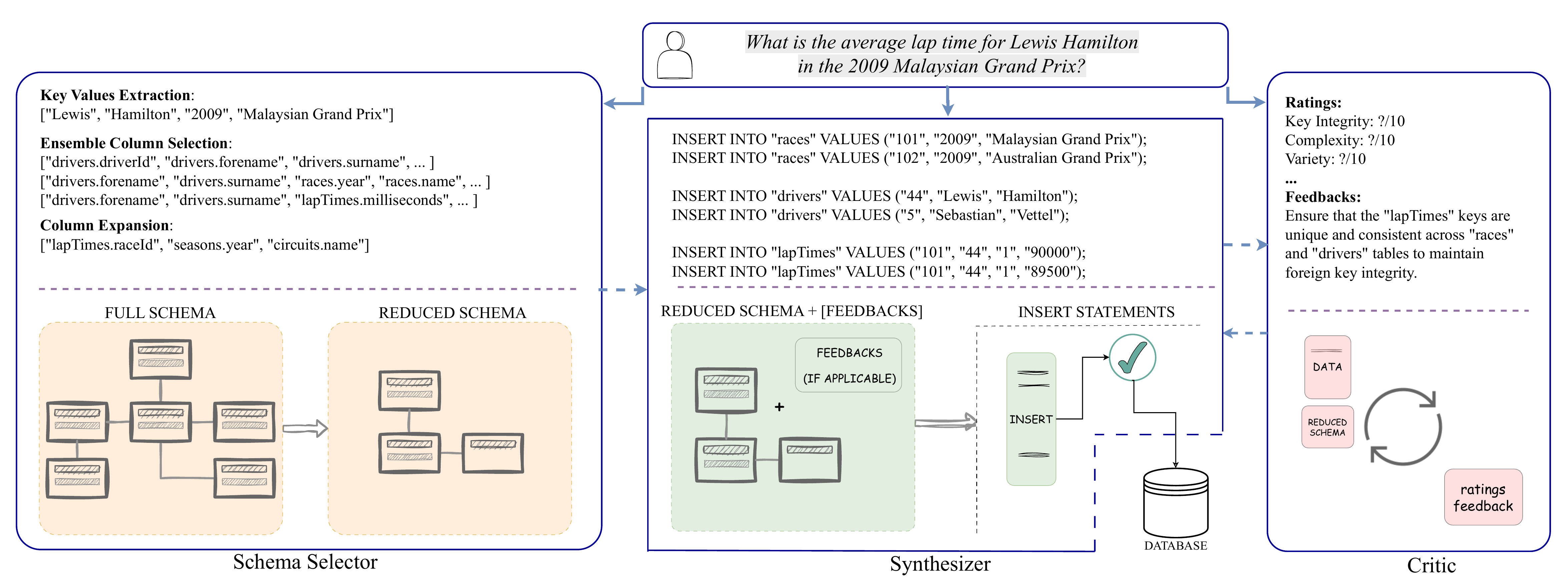}
   \caption{Overview of the SynSQL framework. The schema selector identifies relevant schema elements and reduces the schema space. The synthesizer generates test data based on the NL question and reduced schema. The critic evaluates the quality of the generated data and provides feedback for improvement.}
   \label{fig:overview}
   \end{figure*}
   
   We introduce SynSQL, a modular framework for question-conditioned synthetic database generation. SynSQL decomposes the task into three stages (Figure~\ref{fig:overview}): a \textit{Schema Selector} identifies schema elements relevant to the question, reducing the combinatorial search space; a \textit{Synthesizer} generates relational tuples conditioned on the question and the reduced schema; and a \textit{Critic} enforces structural consistency and semantic alignment, iteratively refining the generated database. This design treats database construction as an interactive structured generation process that integrates language-based reasoning with relational constraints.
   
   We evaluate SynSQL across ten text-to-SQL systems on Spider, BIRD, and Spider~2.0. SynSQL-generated databases reveal consistent robustness gaps: model accuracies drop by 3–14\% compared to evaluation on the original static databases, and in some cases lead to changes in model ranking. This indicates that fixed benchmark instances mask systematic errors. At the same time, SynSQL achieves high constraint satisfaction rates and produces realistic, minimal databases aligned with question semantics. Our analysis highlights both the strengths and limitations of LLMs in structured data synthesis, including failure modes in constraint adherence and semantic grounding under complex schemas.

\paragraph{Contributions.}
This paper makes the following contributions:
(1) We introduce \emph{question-conditioned relational database synthesis} from natural language and schema as a new task for studying the ability of LLMs to construct semantically grounded relational worlds under structural constraints.
(2) We develop \textsc{SynSQL}, a modular framework that decomposes this task into schema selection, question-guided synthesis, and constraint-aware critique with iterative refinement.
(3) Through extensive evaluation across multiple text-to-SQL systems, datasets, and LLM families, we show that SynSQL exposes robustness gaps and induces ranking shifts under structurally valid data variation that are not apparent under static evaluation. 
(4) We provide a detailed analysis of structural validity, semantic support, and failure modes, offering insight into where LLMs succeed and where they remain limited in structured world generation.

   
   
\section{Related Work}
\label{sec:related}

\paragraph{Evaluation and query-centric database generation.}
Text-to-SQL systems are typically evaluated by executing predicted and gold SQL queries on a fixed database instance, as in Spider~\citep{yu2018spider} and BIRD~\citep{li2023can}. While this enables standardized comparison, execution-based evaluation is sensitive to the underlying data and may overestimate correctness due to coincidental agreement on specific instances~\citep{mitsopoulou2025analysis,renggli2025fundamental,zhong2020semantic}. To address this, prior work generates alternative databases that distinguish correct from incorrect queries. Systems such as AGENDA~\citep{deng2005testing}, XData~\citep{veanes2010qex}, and TestSuiteAccuracy~\citep{zhong2020semantic} use query mutation or symbolic analysis, while SMT-based approaches (e.g., VeriEQL~\citep{he2024verieql} and SpotIt~\citep{klopfenstein2025spotit}) synthesize counterexample databases. These methods are inherently \emph{query-centric}, requiring access to gold SQL and reasoning over query structure, which limits scalability and generality.

\paragraph{LLMs for data and structured generation.}
Recent work explores LLMs for synthetic data generation, including controllable generation~\citep{zhou2025difflm} and unified dataset pipelines~\citep{huang2024datagen}. In parallel, LLMs have shown strong capabilities in generating structured outputs such as code, tables, and semi-structured data~\citep{chen2021evaluating,austin2021program,li2023textbooks}, with applications in program synthesis~\citep{chen2021evaluating,austin2021program}, table reasoning~\citep{yin2020tabert,herzig2020tapas}, and tool use~\citep{schick2023toolformer}. Recent work (e.g., StructSynth~\citep{liu2025structsynth}) highlights the difficulty of enforcing structural dependencies in tabular generation. However, these approaches either focus on matching data distributions or generating isolated structured outputs, and do not address relational data generation grounded in natural language and constrained by schema structure.

\paragraph{Our approach.}
We frame relational database construction as a structured generation problem conditioned on natural language and schema. Unlike prior work, we synthesize schema-consistent databases directly from question-schema alignment, without relying on gold queries. This enables evaluation under controlled data variation and provides a testbed for studying how LLMs align language with relational structure under constraints.

\section{Methodology}
\label{sec:method}

We formulate question-conditioned database synthesis as a \emph{structured generation problem} over relational database instances. Given a natural language question $q$ and a database schema $\mathcal{S}$, the goal is to generate a database instance $\mathcal{D}$ such that (i) $\mathcal{D}$ satisfies the structural constraints of $\mathcal{S}$ (e.g., primary and foreign keys), and (ii) $\mathcal{D}$ encodes semantic signals implied by $q$ that enable distinguishing correct and incorrect query interpretations. We introduce \textsc{SynSQL}, a modular framework that decomposes this process into three stages (Figure~\ref{fig:overview}): schema selection, question-conditioned data synthesis, and constraint-aware critique, which together reduce the generation space, construct candidate data, and enforce structural and semantic validity.

\subsection{Schema Selection}
Real-world database schemas often contain many tables and columns, while most NL questions reference only a small subset. 
Providing the full schema can lead to unnecessary or inconsistent data generation (e.g., foreign key violations). The goal is therefore to identify relevant schema elements while preserving sufficient context for coherent synthesis.

Given $\mathcal{S}$ and $q$, we select a subset $\mathcal{S}' \subseteq \mathcal{S}$ by grounding linguistic cues in $q$ to tables, attributes, and relationships. This reduces the combinatorial search space while retaining the structure needed for synthesis.
We implement schema selection using an LLM that identifies relevant elements based on lexical overlap, semantic similarity, and relational context (\S~\ref{sec:prompt-colSel},\ref{sec:prompt-colExp}).

Our approach prioritizes recall, retaining all schema elements that could plausibly support data synthesis. As outlined in Algorithm~\ref{alg:schema-selector} (Appendix), we adopt an ensemble-expansion strategy: the LLM is queried multiple times at different temperature settings to produce diverse candidate subsets, whose union forms a high-recall core. This core is then expanded with semantically or functionally related columns to further improve coverage.

\subsection{Question-Conditioned Data Synthesis}


Unlike prior approaches that rely on reference SQL queries or their mutations, our synthesizer operates directly from $q$ and $\mathcal{S}'$, enabling broader coverage of query semantics. Given $\mathcal{S}'$ and $q$, the synthesizer generates a database instance $\mathcal{D}$ by populating each table with tuples. This requires generating values that are: (i) \textbf{structurally valid}, respecting column types and table schemas, (ii) \textbf{relationally consistent}, maintaining dependencies such as foreign keys, and (iii) \textbf{semantically grounded} in $q$, including values that can expose potential errors in query interpretation.
We leverage the LLM to generate table-by-table tuples conditioned on $q$ and $\mathcal{S}'$, while maintaining consistency between tables through key relationships and value-level dependencies (e.g., foreign keys, aligned attributes and correlated values across tables) (\S~\ref{sec:prompt-synthesizer}).

We apply lightweight postprocessing to enforce basic structural correctness, including dropping tables or columns not in the schema, enforcing arity (padding with \texttt{NULL}s or truncating as needed), and normalizing values based on keywords extracted from the question. This process ensures the database can be loaded and executed, but cannot resolve referential integrity violations or semantic misalignment, which require regenerating coherent tuples and are addressed in the next stage. 

\subsection{Constraint-Aware Critique and Refinement}


Key or referential integrity violations (e.g., foreign keys pointing to missing rows) cannot be fixed by dropping columns; they require the synthesizer to produce a new, consistent set of rows. Similarly, misalignment with question hints, lack of data variety, and oversimplified patterns that inflate success rate without improving discriminative power are semantic issues that only iterative refinement can address. Inspired by self-correction for LLMs~\citep{pan2023automatically}, we therefore introduce a \textbf{Critic} that evaluates each generated database and decides whether to accept it or request a new round of generation. It scores the data on a 1--10 scale across six dimensions: alignment with question hints, key and referential integrity, schema coverage, data complexity, variety in records, and overall relevance.
It translates detected issues into targeted feedback for refinement. If the average score meets the quality threshold (e.g., 8.0), the data is accepted; otherwise, the feedback is incorporated into the next iteration (see Appendix~\ref{sec:appendix-critic} for details).

\paragraph{Discussion:}
This formulation casts database construction as a structured generation problem requiring LLMs to jointly reason over language, schema structure, and relational constraints. Unlike query-centric approaches that derive test data from SQL, SynSQL generates relational instances directly from question--schema alignment, enabling controlled variation of underlying data and more robust evaluation of text-to-SQL systems.





\section{Experimental Evaluation}
Our experimental evaluation aims to address three core questions: (i) Can LLMs generate relational databases that are executable, structurally valid, and semantically supportive? (ii) Do such databases reveal robustness failures not observable under fixed-instance evaluation? (iii) How do SynSQL's components affect generation quality?


\subsection{Experimental Setup}
\paragraph{Datasets.}
We evaluate SynSQL on three widely used text-to-SQL benchmarks: Spider~\citep{yu2018spider}, BIRD~\citep{li2023can}, and Spider 2.0~\citep{lei2024spider}. Spider features simple schemas, while BIRD includes complex queries with joins and nested subqueries. Spider 2.0-SQLite contains 135 enterprise-level problems requiring handling of complex schemas and multi-step queries. This enables comprehensive assessment across varying schema complexity.


\paragraph{Model Configuration.}
We instantiate SynSQL using a mix of proprietary and open-source language models, including GPT-4.1-mini, Gemini-2.5-Flash, Gemini-3-Flash, and Qwen-3-8B. The critic performs up to three refinement iterations and terminates early if the generated database achieves a quality score of 8.0 or higher (on a 10-point scale). 
As a baseline, we use a \emph{vanilla synthesizer} that generates data in a single pass without schema reduction or critic feedback. This baseline is equivalent to SynSQL without schema selection, data validation, or iterative refinement, operating over the full schema with the same prompting strategy.


\paragraph{Text-to-SQL Systems.}
To evaluate the effectiveness of SynSQL-generated databases in distinguishing correct and incorrect queries, we use ten competitive text-to-SQL systems to produce candidate SQL queries:
\textbf{OmniSQL-32B}~\citep{li2025omnisql}, \textbf{RSL-SQL + GPT-4o}~\citep{cao2024rsl}, \textbf{Alpha-SQL + Qwen-32B}~\citep{li2025alpha}, \textbf{CSC-SQL + Qwen-32B}~\citep{sheng2025csc}, \textbf{Gemini-SQL (Multitask SFT + Gemini-2.5-Pro)}~\citep{Pourreza2025GeminiDatabase}, \textbf{DIN-SQL}~\citep{pourreza2023din}, \textbf{DAIL-SQL}~\citep{gao2023text},\textbf{Graphix-3B+PICARD}~\citep{li2023graphix}, \textbf{C3 + ChatGPT}~\citep{dong2023c3}, and \textbf{GPT 5.4}.

\paragraph{Evaluation Metrics.}
We employ three complementary metrics to assess both the quality of the generated databases and their utility for evaluating text-to-SQL systems:

\emph{Success Rate (SR).} SR measures the fraction of questions for which the gold SQL query produces a non-empty result on the generated database.
This indicates whether the synthesized data captures the semantic intent of the natural language question, as expressed by the gold query.
Random or poorly aligned data often yields empty results, making SR a necessary but weak proxy for semantic grounding.


\emph{Execution Accuracy (EX).} 
EX is the standard metric used in Spider and BIRD. It measures the fraction of questions for which the model-generated SQL query produces the same result as the gold SQL query when executed on a single database instance.
EX characterizes model behavior under a fixed database, rather than serving as the primary comparison metric.

\emph{Compound Execution Accuracy (\EXc).}  
\EXc~ extends EX by requiring correctness across both the original benchmark database and the SynSQL-generated database. A prediction is counted as correct only if it 
 produces the same result as the gold SQL query on \emph{both} instances; otherwise, it is considered incorrect. Thus \EXc~ measures robustness under data variation and is upper-bounded  by EX on either database alone.
 

\subsection{Robustness Under Data Variation}
Tables~\ref{tab:results-bird}, \ref{tab:results-spider}, and \ref{tab:spider-2.0-results} show that requiring correctness across both the original and SynSQL-generated database instances leads to consistent and substantial performance drops across all models and datasets. On BIRD, compound execution accuracy (\EXc) decreases by 6-9\% relative to the official execution accuracy, revealing errors that are not exposed under standard single-instance evaluation. Similar trends are observed on Spider and Spider~2.0, confirming that this phenomenon is dataset-independent. 
This gap highlights a fundamental limitation of the commonly-used execution accuracy (EX): it evaluates correctness on a single data instance and can overestimate performance due to coincidental agreement. In contrast, \EXc~ requires correctness across multiple semantically valid database instantiations, providing a stricter and more reliable measure of semantic correctness.





\begin{table}[H]
\centering
\small
\setlength{\tabcolsep}{4pt}
\caption{Drop in execution accuracy ($\Delta$EXc) relative to BIRD (Original). Rankings are shown per row (lower drop is better).}
\label{tab:results-bird}
\begin{tabular}{l|ccccc}
\toprule
\textbf{Method} 
& \textbf{OmniSQL} 
& \textbf{RSL-SQL} 
& \textbf{Alpha-SQL} 
& \textbf{CSC-SQL} 
& \textbf{Gemini-SQL} \\
\midrule

\textbf{BIRD (Original)} 
& 66.75 (5)
& 67.47 (4)
& 69.23 (3)
& 71.06 (2) 
& 72.10 (1) \\ \midrule

Vanilla GPT-4.1-Mini 
& -3.52 (5) 
& -3.39 (4)
& -3.65 (3)
& -3.92 (2) 
& -2.61 (1) \\

Vanilla Gemini-2.5-Flash 
& -3.13 (5) 
& -2.93 (4)
& -3.06 (3)
& -3.52 (2) 
& -2.28 (1) \\

SynSQL (Qwen-3-8B) 
& -7.56 (5)
& -6.39 (4)
& -7.63 (3) 
& -6.70 (2) 
& -5.48 (1) \\

SynSQL (Gemini-2.5-Flash) 
& -6.97 (5)
& -6.42 (4)
& -6.78 (3)
& -7.70 (2) 
& -5.67 (1) \\

SynSQL (GPT-4.1-Mini) 
& \textbf{-8.21 (5)}
& \textbf{-7.43 (4)}
& \textbf{-8.80 (3)}
& \textbf{-9.07 (2)} 
& \textbf{-6.34 (1)} \\

\bottomrule
\end{tabular}
\end{table}

\begin{table}[H]
\centering
\small
\setlength{\tabcolsep}{4pt}
\caption{Drop in execution accuracy ($\Delta$EXc) relative to Spider (Original). Rankings are shown per row.}
\label{tab:results-spider}
\begin{tabular}{l|cccc}
\toprule
\textbf{Method} 
& \textbf{Graphix} 
& \textbf{C3} 
& \textbf{DIN-SQL} 
& \textbf{DAIL-SQL} \\
\midrule

\textbf{Spider (Original)} 
& 77.51 (4) 
& 77.76 (3) 
& 80.46 (2) 
& 80.66 (1) \\ \midrule

Vanilla GPT-4.1-Mini
& -2.36 (3) 
& -3.58 (4) 
& -2.48 (2) 
& -2.03 (1) \\

Vanilla Gemini-2.5-Flash
& -2.46 (3) 
& -2.52 (4) 
& -2.80 (2) 
& -1.94 (1) \\

SynSQL (Qwen-3-8B)
& -3.14 (3) 
& -6.10 (4) 
& -3.44 (2) 
& -3.00 (1) \\

SynSQL (Gemini-2.5-Flash)
& -2.95 (3) 
& -5.13 (4) 
& -3.19 (2) 
& -2.42 (1) \\

SynSQL (GPT-4.1-Mini)
& \textbf{-3.14 (3)} 
& \textbf{-5.71 (4)} 
& \textbf{-4.15 (2)} 
& \textbf{-3.10 (1)} \\

\bottomrule
\end{tabular}
\end{table}

\begin{wraptable}{r}{0.5\textwidth}
\vspace{-2em}
\begin{minipage}{0.46\textwidth}
\centering
\caption{Drop in execution accuracy ($\Delta$EXc) relative to Spider~2.0-SQLite (Original).}
\label{tab:spider-2.0-results}
\vspace{0.15em}
\begingroup
\footnotesize
\setlength{\tabcolsep}{2.5pt}
\renewcommand{\arraystretch}{1.05}
\begin{tabular}{@{}>{\raggedright\arraybackslash}p{0.58\linewidth}|c|c@{}}
\toprule
\textbf{Method} 
& \textbf{OmniSQL} 
& \textbf{GPT 5.4} \\
\midrule

\textbf{Spider 2.0 (Original)} 
& 12.59 (2)
& 40.00 (1) \\ \midrule

Vanilla GPT-4.1-Mini 
& -1.48 (2)
& -8.15 (1)\\

Vanilla Gemini-3-Flash 
& -3.7 (2)
& -12.59 (1)\\

SynSQL (GPT-4.1-Mini) 
& -2.97 (2) 
& -10.37 (1) \\

SynSQL (Gemini-3-Flash) 
& \textbf{-5.18 (2)} 
& \textbf{-14.07 (1)} \\

\bottomrule
\end{tabular}
\endgroup
\end{minipage}
\end{wraptable}

\paragraph{Performance Gap and Model ranking.}
Model rankings remain largely stable on BIRD and Spider~2.0, although we observe significant changes in performance gaps between models.
However, on Spider even the ranking among models changes. This suggests that performance and ranking stability is subject to change, with data variation having a greater impact when competing models exhibit similar performance on a single database instance. In such cases, evaluating across multiple database instances reveals finer-grained differences that are otherwise hidden. Under SynSQL, Graphix is ranked above C3 on Spider, reversing their order under the official evaluation and exposing errors masked by single-instance testing.

\subsection{Quality of Synthesized Databases}
SynSQL can also serve as a standalone data synthesizer when human-curated databases are not available. We therefore evaluate the quality of SynSQL-generated data by comparing it against both human-curated databases and vanilla LLM-based baselines.

\paragraph{Success Rate.}
As shown in Table~\ref{tab:results-ex} (Appendix), SynSQL achieves success rates of 82.07\% on BIRD, 93.04\% on Spider, and 80.0\% on Spider~2.0, consistently outperforming vanilla baselines across all datasets and LLM families. Notably, on Spider, SynSQL even surpasses the original human-authored database (92.55\%) when paired with Gemini-2.5-Flash and GPT-4.1-Mini. 
This improvement can be attributed to inconsistencies in the benchmark data, such as missing values and formatting artifacts (e.g., trailing spaces), which can lead to unintended empty results during execution. By synthesizing clean and semantically aligned data, SynSQL mitigates these issues and better reflects the intent of the underlying queries. Illustrative examples of such inconsistencies are provided in Figures~\ref{fig:sr-example-6} and~\ref{fig:sr-example-7} in the Appendix.

\paragraph{Execution Accuracy.}
We further evaluate execution accuracy (EX) on each method’s generated database (Table~\ref{tab:results-ex} in the Appendix). SynSQL yields EX values that are close to those on the original human-curated databases across BIRD, Spider, and Spider~2.0, while remaining consistently lower than vanilla baselines. Here, lower EX indicates a more discriminative evaluation setting. 
We observe higher EX for vanilla baselines, particularly on BIRD and Spider, indicating that their generated data fails to effectively discriminate between correct and incorrect queries.
In contrast, SynSQL maintains high SR while producing more demanding databases, resulting in lower but more informative EX. Importantly, these differences are not due to invalid data: SynSQL-generated databases remain executable, structurally valid, and semantically supportive. Overall, this demonstrates that SynSQL enables more faithful and rigorous execution-based evaluation.

\paragraph{Structural Validity.}
Figure~\ref{fig:relational-validity} (Appendix) presents the percentage of generated databases that adhere to schema constraints, including primary and foreign key integrity and table structure, are executable and have valid data. SynSQL achieves near-perfect validity (99\% across all datasets), improving over vanilla LLM baselines by 30--33\% on BIRD, 9--20\% on Spider, and 18--25\% on Spider~2.0. The gains are most pronounced on schema-complex datasets such as BIRD and Spider~2.0, highlighting the difficulty of maintaining inter-table dependencies without explicit schema-aware guidance. These results demonstrate that, with schema reduction and iterative refinement, LLMs can reliably generate data that respects both structural constraints and inter-table dependencies. Combined with strong SR and EX performance, this confirms that SynSQL produces databases that are not only semantically meaningful but also structurally sound.

\subsection{Failure Analysis}
\label{sec:failure-analysis-sr}

Despite strong overall performance, a systematic analysis reveals recurring failure modes that expose fundamental limitations of LLM-based structured data generation.

\begin{figure}[t]
   \centering
   \includegraphics[width=1\textwidth]{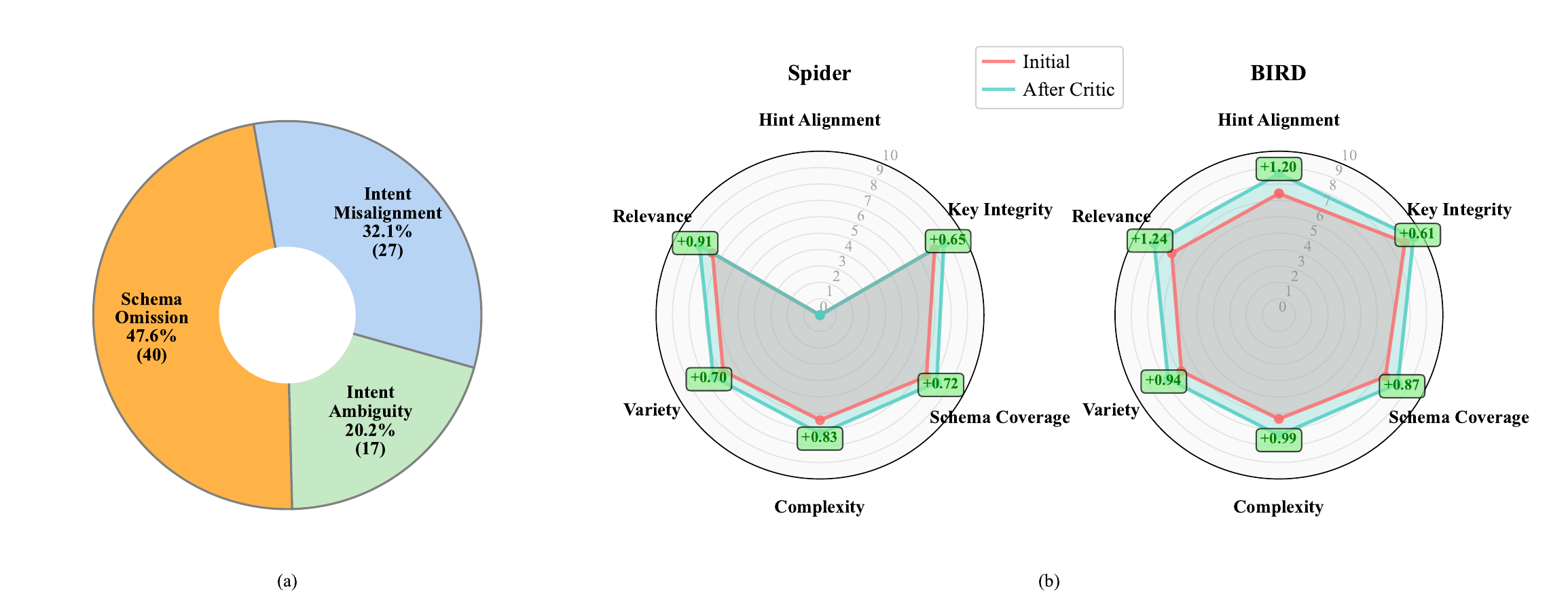}
   \vspace{-0.3em}
   \caption{\small \textbf{(a)} Breakdown of failure cases (84 failures out of 500 BIRD dev questions). Schema selector failures: omitted tables/columns. Semantic failures: alignment mismatches and NL ambiguity. \textbf{(b)} Impact of the critic component on each of the five data quality criteria in SynSQL, using Gemini-2.5-Flash on Spider and BIRD dev sets. Spider results exclude Hint Alignment as evidence/hint entries are not present in Spider.}
   \label{fig:sr-pie-chart}
\end{figure}

To understand where synthesis fails to capture the structural or semantic cues of the NL question, we analyzed success rate failures on a random sample of 500 BIRD questions. Of these, 84 cases yield empty results when executing the gold query on the SynSQL-generated database. Figure~\ref{fig:sr-pie-chart}(a) summarizes the breakdown of these failures and highlights two primary sources of error.

\paragraph{Adherence to Schema Constraints.} In these cases, the schema selection omits one or more tables or columns required by the gold query, preventing the synthesized database from supporting the intent of the question. For example, if the gold query references a \texttt{player} table but the schema selector fails to include it in the reduced schema, the generated database will lack the necessary structure to yield a non-empty result. These failures highlight the difficulty of high-recall schema selection under complex schemas with many tables and columns.

\paragraph{Semantic Failures.} Here the generated data is structurally valid and executable, but the instantiated values do not align with the expectations of the gold query. Common issues are case mismatches (e.g., \texttt{owner} vs.\ \texttt{OWNER}), differences in value granularity (e.g., \texttt{Prague 1} vs.\ \texttt{Prague}), and inconsistencies with benchmark-specific conventions. Among the 44 semantic failures, 27 are attributable to SynSQL generation, while the remaining 17 stem from ambiguous or inconsistent question--query pairs in the BIRD dev set. These cases illustrate the difficulty of grounding generated values to a single ``correct'' interpretation when the question or gold query is underspecified.

Overall, these failure modes reveal two key challenges in question-conditioned database synthesis: (i) achieving high-recall schema selection without access to gold queries, and (ii) ensuring consistent semantic grounding of values (e.g., case, format, granularity) under database-specific conventions. 
While LLMs can generate plausible relational data, precise semantic control and constraint adherence under complex schemas remain challenging. Improving controllability and constraint-aware generation is an important direction for future work.
For detailed examples and further discussion, see Appendix~\ref{sec:appendix-error-analysis}.

\subsection{Ablation Studies}
\label{sec:ablation}
We conduct ablation studies to quantify the contribution of key components in SynSQL and to better understand the trade-offs underlying its design.

\paragraph{Effect of the critic.}
\begin{wrapfigure}{r}{0.55\textwidth}
\centering
\begin{tikzpicture}
\begin{axis}[
   ybar,
   bar width=10pt,
   width=0.48\textwidth,
   height=4.2cm,
   enlarge x limits=0.18,
   ylabel={Success Rate (\%)},
   ymin=65, ymax=85,
   axis y line*=left,
   xtick=data,
   symbolic x coords={
      GPT-4.1-Mini, GPT-4.1-Mini (w/o Critic),
      Gemini-2.5-Flash, Gemini-2.5-Flash (w/o Critic),
      Qwen-3-8B, Qwen-3-8B (w/o Critic)
   },
   xticklabel style={rotate=25, anchor=east, font=\scriptsize},
   nodes near coords,
   nodes near coords style={font=\tiny, yshift=2pt, color=black},
   bar shift=0pt,
   every node near coord/.append style={/pgf/number format/precision=2},
   tick label style={font=\scriptsize},
   major grid style={dashed,draw=gray!30},
   grid=major,
   legend style={at={(0.5,-0.18)}, anchor=north, legend columns=-1, font=\scriptsize},
   ]
\addplot[fill=teal!70!white] coordinates {
   (GPT-4.1-Mini,82.07) (GPT-4.1-Mini (w/o Critic),82.00)
   (Gemini-2.5-Flash,0) (Gemini-2.5-Flash (w/o Critic),0)
   (Qwen-3-8B,0) (Qwen-3-8B (w/o Critic),0)
};
\addplot[fill=violet!70!white] coordinates {
   (GPT-4.1-Mini,0) (GPT-4.1-Mini (w/o Critic),0)
   (Gemini-2.5-Flash,80.57) (Gemini-2.5-Flash (w/o Critic),78.03)
   (Qwen-3-8B,0) (Qwen-3-8B (w/o Critic),0)
};
\addplot[fill=orange!80!white] coordinates {
   (GPT-4.1-Mini,0) (GPT-4.1-Mini (w/o Critic),0)
   (Gemini-2.5-Flash,0) (Gemini-2.5-Flash (w/o Critic),0)
   (Qwen-3-8B,73.60) (Qwen-3-8B (w/o Critic),67.86)
};
\end{axis}
\end{tikzpicture}
\caption{Impact of the critic component on success rate (\%) of SynSQL with three different LLMs on the BIRD dev set.}
\label{fig:critic-ablation}
\end{wrapfigure}
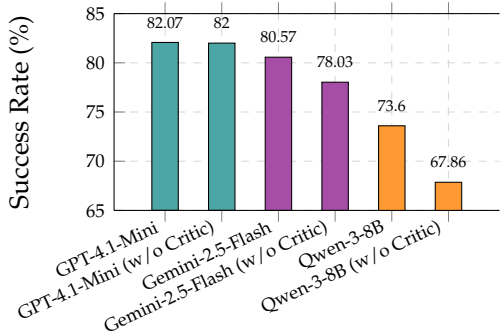

As illustrated in Figure~\ref{fig:critic-ablation}, the critic component has a consistent and often substantial impact on success rate across LLMs on the BIRD dev set. Incorporating the critic improves performance for all models, with the largest gain observed for Qwen-3-8B (from 67.86\% to 73.60\%). The improvement is more modest for GPT-4.1-Mini, which already performs strongly, but the critic still contributes to more stable and reliable outcomes.
These results highlight the critic’s role in refining synthetic data through iterative feedback. By enforcing alignment with question intent, schema constraints, and data diversity, the critic guides the synthesizer toward more accurate and executable outputs. This effect is particularly pronounced for smaller or less capable models, which benefit more from corrective feedback.

To further analyze this effect, Figure~\ref{fig:sr-pie-chart}(b) breaks down the critic’s contributions across its evaluation criteria, including hint alignment, key integrity, schema coverage, data complexity, data variety, and relevance. Improvements are observed across all dimensions on both Spider and BIRD, indicating that the critic enhances both semantic coherence and diversity of the generated databases. Additional results on compound execution accuracy (\EXc) and qualitative examples of critic feedback are provided in Appendix~\ref{sec:appendix-critic}.

\paragraph{Effect of Schema Selection.}
Table~\ref{tab:abstudy} shows the effect of schema selection on performance. The SynSQL method with ensemble-expansion outperforms all ablated versions, confirming that both phases contribute meaningfully to success rate, especially on BIRD, where complex schemas increase the difficulty of accurate column selection. Using the oracle schema yields highest success rate, indicating further improvements in schema selection could enhance performance.
We also observe that the average column count selected by the schema selector is significantly lower than the full schema, demonstrating SynSQL's ability to generate compact databases while maintaining high success rates.
However, aggressive reduction risks omitting columns required by gold queries, causing otherwise correct queries to fail. This highlights the inherent tension between minimizing schema complexity and preserving query executability. Despite this limitation, SynSQL's ensemble-expansion strategy achieves a balance that maintains high success rates while generating significantly more compact databases than the full schema.

As illustrated in Appendix Figures~\ref{fig:db-example-1} and \ref{fig:db-example-2}, these compact synthetic databases are easier to inspect and validate, facilitating future directions such as human-in-the-loop evaluation and generating expected outputs via table reasoning.

\begin{table}[ht]
\centering
\footnotesize
\caption{\label{tab:abstudy}Ablation study on schema selection in SynSQL. We report success rate (SR) and average column count (CC) selected by the schema selector on the BIRD and Spider dev sets. All variants use GPT-4.1-Mini as both the base model and the critic. \textbf{SynSQL w Oracle Schema} assumes perfect schema recall and serves as an upper bound on performance under ideal schema selection.}
\begin{tabular}{l|cc|cc}
\toprule
& \multicolumn{2}{c|}{BIRD} & \multicolumn{2}{c}{Spider} \\
Method & SR (\%) & CC & SR (\%) & CC \\
\midrule
SynSQL w Oracle Schema & 91.46 & 4.71 & 94.58 & 2.85 \\
\arrayrulecolor{gray!40}\midrule\arrayrulecolor{black}
SynSQL & 82.07 & 8.37 & 93.04 & 6.71 \\
SynSQL w/o Expansion & 79.53 & 5.42 & 92.75 & 3.92 \\
SynSQL w/o Ensemble-Expansion & 77.38 & 4.99 & 91.88 & 3.56 \\
SynSQL w/o Schema Selection (Full-Schema) & 71.25 & 75.56 & 92.94 & 24.55 \\
\bottomrule
\end{tabular}
\end{table}


\section{Conclusion}
We introduced \textsc{SynSQL}, a framework for synthesizing relational databases from natural language questions and schema, and showed that database generation provides a powerful lens for evaluating text-to-SQL systems. Our results demonstrate that standard single-instance evaluation substantially overestimates performance: when evaluated across structurally valid database variations, all models exhibit consistent drops and, in some cases, changes in ranking.
Beyond evaluation, our findings highlight both the capabilities and limitations of LLMs in structured generation. While LLMs can produce executable, structurally valid, and semantically aligned relational data, failures in schema adherence and semantic grounding reveal challenges in controllability, particularly under complex schemas and ambiguous query intent. This positions relational data synthesis as a useful testbed for studying structured reasoning in LLMs.
Overall, question-conditioned database synthesis offers a new framework for studying evaluation, controllability, and robustness in large language models.

Our work suggests several directions for future research. First, synthesis without gold queries enables \emph{test-time evaluation}: in real-world settings without annotations, model-generated queries can be validated against synthesized databases to ensure semantic support. Developing stronger evaluation criteria, e.g., consistency across diverse synthesized databases or adversarial data generation, is an important next step. Second, improving controllability remains a key challenge, as does extending synthesis to more complex settings such as multi-database scenarios and interactive workflows.

\bibliography{colm2026_conference}
\bibliographystyle{colm2026_conference}

\appendix
\section{Appendix}

\subsection{Schema Selection Algorithm}
\begin{algorithm}[H]
\caption{Schema Selector for SynSQL}
\label{alg:schema-selector}
\begin{algorithmic}[1]
\Require{Natural language question $Q$, full schema $S$, auxiliary knowledge $K$, model $\pi$}
\State\textbf{Ensure:} Reduced schema $S_{reduced}$ relevant to $Q$
\State Initialize $S_{core} \gets \emptyset$
\For{temperature $t$ in $\{0, 0.3, 0.7\}$}
   \State Query $\pi$ with $(Q, S, K)$ at temperature $t$ to extract core elements
   \State $S_{core} \gets S_{core} \cup$ elements returned by $\pi$
\EndFor
\State Query $\pi$ for semantically related columns to $S_{core}$
\State $S_{aux} \gets$ related columns returned by $\pi$
\State $S_{reduced} \gets S_{core} \cup S_{aux}$
\State\textbf{return} $S_{reduced}$
\end{algorithmic}
\end{algorithm}

\subsection{Execution Accuracy (EX) Details}
\label{sec:appendix-execution-accuracy}
The main results table (Tables~\ref{tab:results-bird}, \ref{tab:results-spider}, and \ref{tab:spider-2.0-results}) report compound execution accuracy (\EXc) on BIRD, Spider, and Spider~2.0. Here we provide the full normal execution accuracy (EX) breakdown for all three.

\begin{table}[H]
\centering
\small
\setlength{\tabcolsep}{3pt}
\caption{\label{tab:results-ex}Execution accuracy (EX) on BIRD (a), Spider (b), and Spider~2.0-SQLite (c). SR: success rate (\%); EX: execution accuracy measured on each method's database.}
\vspace{0.3em}
\begin{center}
\textbf{(a) BIRD dev set.}\\[0.5em]
\begin{tabular}{@{}l@{\hspace{4pt}}c@{\hspace{7pt}}|ccccc@{}}
\toprule
\textbf{Method} & \textbf{SR} & \textbf{OmniSQL} & \textbf{RSL-SQL} & \textbf{Alpha-SQL} & \textbf{CSC-SQL} & \textbf{Gemini-SQL} \\
\cmidrule(lr){3-7}
& & EX & EX & EX & EX & EX \\
\midrule
BIRD (Original) & 99.87 & 66.75 & 67.47 & 69.23 & 71.06 & 72.10 \\
Vanilla GPT-4.1-Mini & 69.43 & 80.51 & 81.68 & 82.00 & 82.20 & 84.68 \\
Vanilla Gemini-2.5-Flash & 67.14 & 79.53 & 80.31 & 78.42 & 80.64 & 82.86 \\
SynSQL (Qwen-3-8B) & 73.60 & 68.84 & 70.01 &68.45 & 71.19 & 74.64 \\
SynSQL (Gemini-2.5-Flash) & 80.57 & 68.32 & 69.03 & 66.69 & 70.08 & 73.21 \\
SynSQL (GPT-4.1-Mini) & \textbf{82.07} & 65.71 & 67.54 & 67.20 &68.12 & 73.14 \\
\bottomrule
\end{tabular}
\end{center}

\vspace{1.2em}
\begin{center}
\textbf{(b) Spider dev set.}\\[0.5em]
\begin{tabular}{@{}l@{\hspace{4pt}}c@{\hspace{10pt}}|cccc@{}}
\toprule
\textbf{Method} & \textbf{SR} & \textbf{Graphix} & \textbf{C3} & \textbf{DIN-SQL} & \textbf{DAIL-SQL} \\
\cmidrule(lr){3-6}
& & EX & EX & EX & EX \\
\midrule
Spider (Original) & 92.55 & 77.51 & 77.76 & 80.46 & 80.66 \\
Vanilla GPT-4.1-Mini & 91.88 & 79.88 & 79.78 & 82.85 & 83.07 \\
Vanilla Gemini-2.5-Flash & 82.59 & 81.14 & 82.79 & 83.08 & 84.24 \\
SynSQL (Qwen-3-8B) & 77.18 & 80.95 & 79.50 & 81.72 & 84.24 \\
SynSQL (Gemini-2.5-Flash) & 92.84 & 78.33 & 77.47 & 81.33 & 81.72 \\
SynSQL (GPT-4.1-Mini) & \textbf{93.04} & 77.95 & 76.02 & 79.79 & 81.24 \\
\bottomrule
\end{tabular}
\end{center}

\vspace{1.2em}
\begin{center}
\textbf{(c) Spider~2.0-SQLite.}\\[0.5em]
\begin{tabular}{@{}l@{\hspace{4pt}}c@{\hspace{8pt}}|cc@{}}
\toprule
\textbf{Method} & \textbf{SR} & \textbf{OmniSQL} & \textbf{GPT 5} \\
\cmidrule(lr){3-4}
& & EX & EX \\
\midrule
Spider 2.0 (Original) & 92.55 & 12.59 & 40.00 \\
Vanilla GPT-4.1-Mini & 63.70 & 31.85 & 53.33 \\
Vanilla Gemini-3-Flash & 75.56 & 25.19 & 50.37 \\
SynSQL (GPT-4.1-Mini) & 68.89 & 22.22 & 48.89 \\
SynSQL (Gemini-3-Flash) & \textbf{80.00} & 17.78 & 45.93 \\
\bottomrule
\end{tabular}
\end{center}
\end{table}

\subsection{Relational Validity and Data Completeness}

\begin{figure}[H]
    \centering
    \includegraphics[width=1\textwidth]{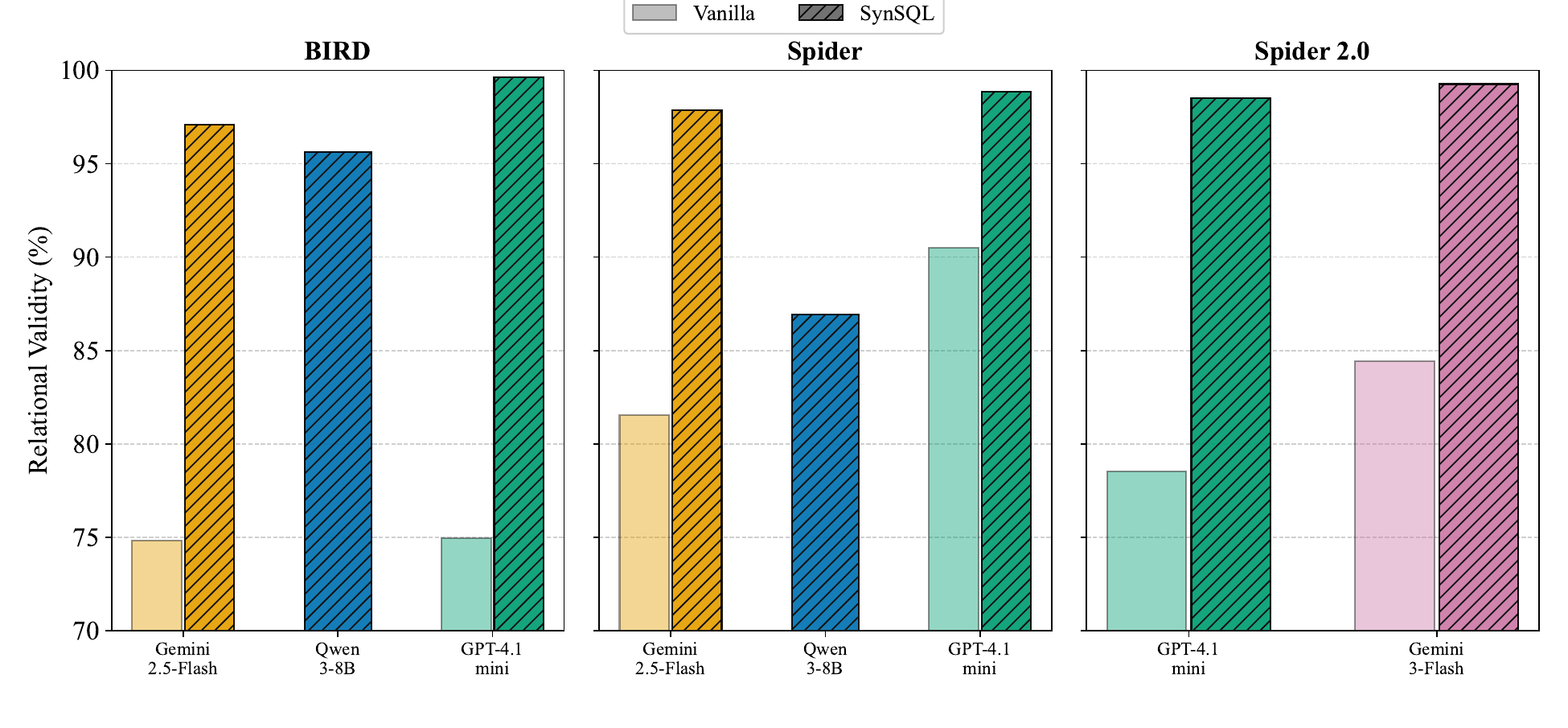} 
    \caption{Relational validity and data completeness of SynSQL and Vanilla-generated databases.}
    \label{fig:relational-validity}
\end{figure}

\subsection{Success Rates and Compound Execution Accuracy by Schema Complexity}
We define schema complexity levels based on the number of columns involved in the gold SQL query. Specifically, we count the total number of columns across all tables referenced in each gold query; higher column counts generally correlate with more complex joins, filters, and reasoning steps.
Based on the distribution of complexity levels in the BIRD dataset, we define three buckets:
\textbf{Low Complexity:} Questions with a total column count of 1-15.
\textbf{Medium Complexity:} Questions with a total column count of 16-60.
\textbf{High Complexity:} Questions with a total column count of 61 or more.

\begin{figure}[H]
    \centering
    \includegraphics[width=1\textwidth]{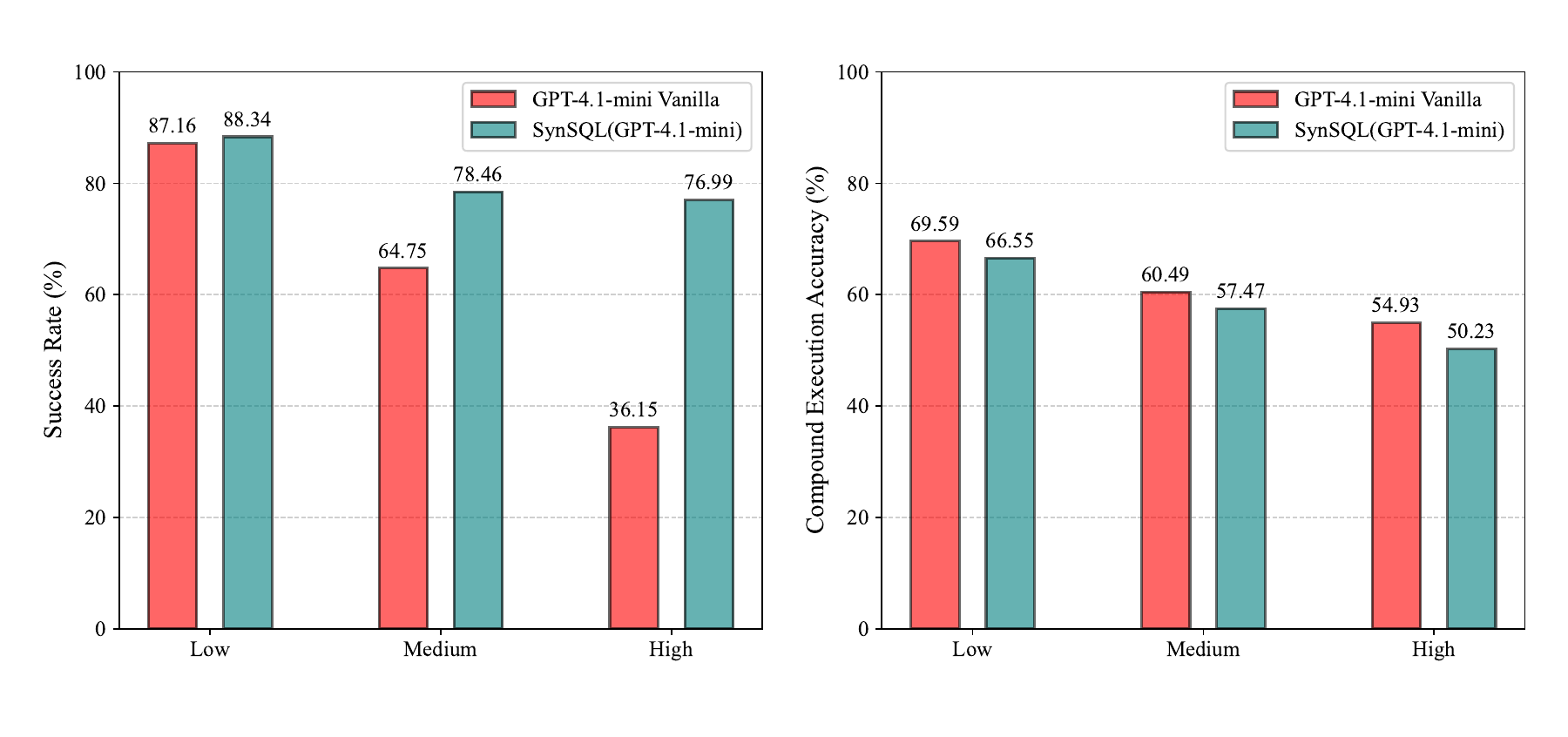} 
    \caption{Success  rate and compound execution accuracy of SynSQL vs.\ GPT-4.1-Mini on BIRD dev set, broken down by schema complexity (Low: 1-15 columns, Medium: 16-60 columns, High: 61+ columns).}
    \label{fig:success-rate-difficulty}
\end{figure}

As shown in Figure~\ref{fig:success-rate-difficulty}, SynSQL consistently outperforms vanilla GPT-4.1-mini. The performance gap in success rate widens as schema complexity increases, and SynSQL’s execution accuracy remains consistently lower than that of vanilla GPT-4.1-mini.

\subsection{Additional Error Analysis and Examples (Success Rate)}
\label{sec:appendix-error-analysis}

This section provides detailed examples and extended discussion of the success-rate failure modes summarized in Section~\ref{sec:failure-analysis-sr} (Figure~\ref{fig:sr-pie-chart}). Of the 84 failed questions, 40 were due to schema selector failures. In these cases, schema reduction led to the omission of one or more tables or columns used in the gold query.
This does not necessarily mean the generated data is meaningless; rather, the human annotator who wrote the gold query may have targeted different schema elements than the LLM. For example, in question 387 from the \texttt{card\_games} database (Figure~\ref{fig:sr-example-1}):

\begin{figure}[H]
    \centering
    \includegraphics[width=1\textwidth]{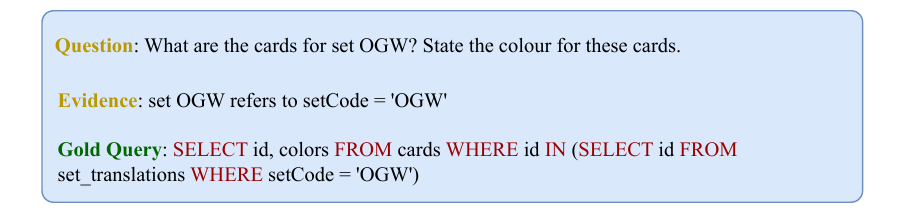} 
    \vspace{-1.5em}
    \caption{\small An example of schema selection failure. The synthetic data omits the \texttt{setCode} column from \texttt{set\_translations}, leading to a failed query.}
    \label{fig:sr-example-1}
\end{figure}

SynSQL has generated data for the \texttt{setCode} column in \texttt{cards}, but omitted the \texttt{setCode} column from \texttt{set\_translations} during schema selection.
The gold query joins both tables on \texttt{setCode}, leading to failure. However, the synthetic data still contains valid \texttt{setCode} values, just not in the joined table. This highlights the challenge of schema selection in open-ended text-to-SQL tasks, where multiple valid interpretations exist.

The remaining 44 failures were due to misinterpretation of question intent. For example, in question 156 from the financial database (Figure~\ref{fig:sr-example-2}):

\begin{figure}[H]
   \centering
   \includegraphics[width=1\textwidth]{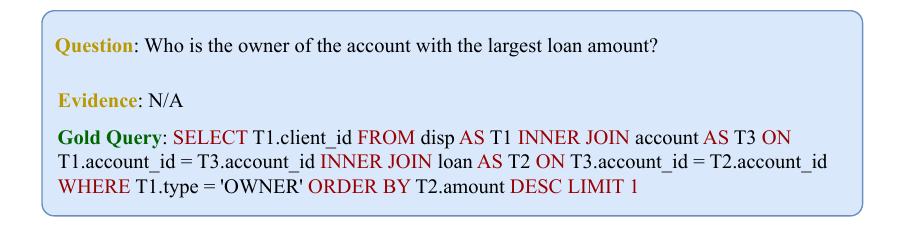} 
   \vspace{-1.5em}
   \caption{\small Example of misinterpretation: the synthetic data contains values such as \texttt{owner} (lowercase) in the \texttt{type} column, while the gold query expects \texttt{OWNER} (uppercase). This case sensitivity mismatch leads to a failed query.}
   \label{fig:sr-example-2}
\end{figure}

Here, the synthetic database reflects the casing found in the question or evidence, but the gold query expects a different case. Such mismatches between generated data and gold query expectations, especially regarding case sensitivity or value formatting, can result in lower success rates.

Another example is question 90 from the financial database (Figure~\ref{fig:sr-example-3}):

\begin{figure}[H]
   \centering
   \includegraphics[width=1\textwidth]{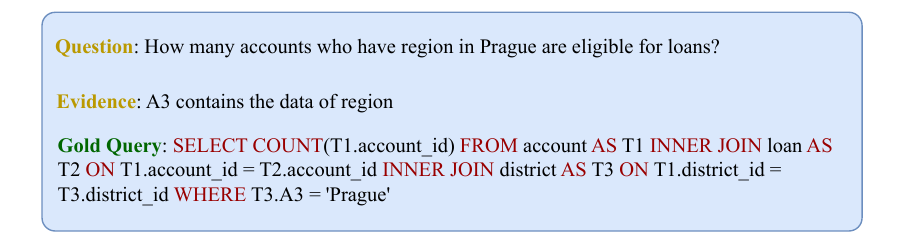} 
   \vspace{-1.5em} 
   \caption{\small An example of misinterpretation due to synthetic data not matching gold query conditions. The synthetic data contains values that do not satisfy the gold query's WHERE clause, leading to failure.}
   \label{fig:sr-example-3}
\end{figure}

The gold query expects \texttt{district.A3 = 'Prague'}, but the synthetic data contains values such as \texttt{Prague 1}, \texttt{Prague 2}, and \texttt{Prague 3}. Here, the LLM generated region names with appended numbers, resulting in a mismatch with the gold query's expected value.

Some misinterpretations are due to misalignment between the question and the gold query in the BIRD dev set, rather than errors by SynSQL. For example, in question 803 from the \texttt{Superhero} database (Figure~\ref{fig:sr-example-4}):

\begin{figure}[H]
   \centering
   \includegraphics[width=1\textwidth]{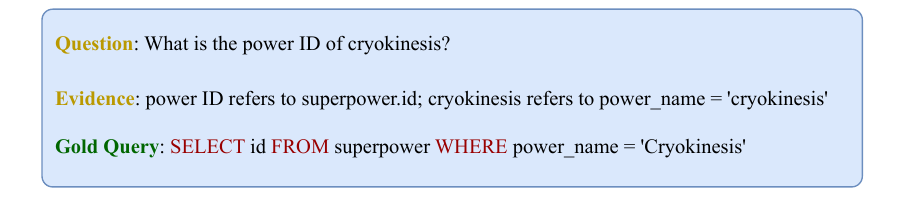} 
   \vspace{-1.5em} 
   \caption{\small An example of misinterpretation due to inconsistencies between question/evidence and gold query in the BIRD dev set. The synthetic data aligns with the question, but not the gold query, leading to failure.}
   \label{fig:sr-example-4}
\end{figure}

In this case, the question and evidence refer to \texttt{cryokinesis} (lowercase), while the gold query expects \texttt{'Cryokinesis'} (capitalized). The synthetic database contains \texttt{power\_name = 'cryokinesis'}, resulting in a mismatch with the gold query and subsequently lower success rate.
Similarly, in question 758, the question and evidence specify \texttt{race = 'human'}, but the gold query expects \texttt{race = 'Human'}. In question 415, the question and evidence use \texttt{Status = 'legal'}, while the gold query expects \texttt{Status = 'Legal'}. The synthetic data generated by SynSQL reflects the casing found in the question, leading to mismatches with the gold query.

In summary, among the 44 misinterpretation cases, 17 stem from insufficient or ambiguous information in the BIRD dev set, while 27 are attributable to SynSQL's generation errors. The following BIRD dev set questions could not be correctly handled by SynSQL due to a lack of necessary information in the dataset for generating appropriate synthetic data. Such cases are likely to be challenging for any text-to-SQL system: 22, 73, 180, 309, 415, 758, 769, 803, 815, 818, 871, 1194, 1336, 1472, 1491, 1499, and 1528.

\subsection{Limitations}

SynSQL demonstrates strong performance in generating synthetic databases for text-to-SQL evaluation, but it has limitations.
The schema selection process may omit relevant tables or columns, leading to gold queries returning empty results.
This remains an active area of research in text-to-SQL evaluation. The challenge is amplified in our data synthesis setting, where the absence of actual database contents and value-based retrieval mechanisms makes high-recall schema selection inherently difficult. However, several practical extensions could improve robustness while maintaining the minimalist design principle. First, implementing multi-hop schema traversal guided by LLMs could recover essential join paths and connector tables in complex schemas, adding minimal columns while significantly boosting recall. Second, employing ensemble methods across multiple LLMs could reduce interpretation variance and yield more stable column predictions. These approaches offer promising directions for addressing the remaining failure cases while preserving SynSQL's core advantages.

Additionally, SynSQL relies on the assumptions made by the large language models used. If the LLMs misinterpret the question intent or generate inconsistent data, this can lead to lower success rates. Incorporating additional constraints or validation steps during data generation could help mitigate this.

\subsection{Analysis on Effect of Critic Component}
\label{sec:appendix-critic}

The main body (Section~\ref{sec:failure-analysis-sr}, Figure~\ref{fig:sr-pie-chart}(b)) describes the critic's impact across the six data quality criteria. This section provides additional analysis and examples: compound execution accuracy ablations, the critic's role in avoiding oversimplified data patterns, and representative critic feedback.

As shown in Figure~\ref{fig:critic-ablation}, the critic component consistently improves SynSQL's success rate across different LLMs on the BIRD dev set. The spider charts in Figure~\ref{fig:sr-pie-chart}(a) (main body) illustrate the same breakdown by criteria.

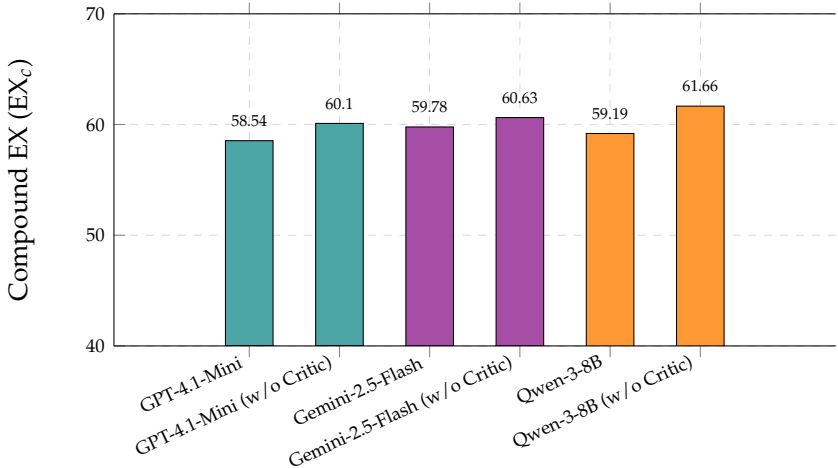
\begin{figure}[h]
\centering
\begin{tikzpicture}
\begin{axis}[
   ybar,
   bar width=18pt,
   width=0.8\textwidth,
   height=6cm,
   enlarge x limits=0.3,
   ylabel={Compound EX (\EXc)},
   ymin=40, ymax=70,
   axis y line*=left,
   xtick=data,
   symbolic x coords={
      GPT-4.1-Mini, GPT-4.1-Mini (w/o Critic),
      Gemini-2.5-Flash, Gemini-2.5-Flash (w/o Critic),
      Qwen-3-8B, Qwen-3-8B (w/o Critic)
   },
   xticklabel style={rotate=25, anchor=east, font=\scriptsize},
   nodes near coords,
   nodes near coords style={font=\tiny, yshift=2pt, color=black},
   bar shift=0pt,
   every node near coord/.append style={/pgf/number format/precision=2},
   tick label style={font=\scriptsize},
   major grid style={dashed,draw=gray!30},
   grid=major,
   legend style={at={(0.5,-0.18)}, anchor=north, legend columns=-1, font=\scriptsize},
   ]
\addplot[fill=teal!70!white] coordinates {
   (GPT-4.1-Mini,58.54) (GPT-4.1-Mini (w/o Critic),60.10)
   (Gemini-2.5-Flash,0) (Gemini-2.5-Flash (w/o Critic),0)
   (Qwen-3-8B,0) (Qwen-3-8B (w/o Critic),0)
};
\addplot[fill=violet!70!white] coordinates {
   (GPT-4.1-Mini,0) (GPT-4.1-Mini (w/o Critic),0)
   (Gemini-2.5-Flash,59.78) (Gemini-2.5-Flash (w/o Critic),60.63)
   (Qwen-3-8B,0) (Qwen-3-8B (w/o Critic),0)
};
\addplot[fill=orange!80!white] coordinates {
   (GPT-4.1-Mini,0) (GPT-4.1-Mini (w/o Critic),0)
   (Gemini-2.5-Flash,0) (Gemini-2.5-Flash (w/o Critic),0)
   (Qwen-3-8B,59.19) (Qwen-3-8B (w/o Critic),61.66)
};
\end{axis}
\end{tikzpicture}
\caption{Impact of the critic component on compound execution accuracy (\EXc) of SynSQL for OmniSQL with three different LLMs on the BIRD dev set. Showing that the critic improves the database's ability to distinguish between correct and incorrect SQL queries.}
\label{fig:critic-exc}
\end{figure}

\paragraph{Critic's Role in Avoiding Oversimplified Data Patterns}
One potential limitation of LLM-based synthesis is the tendency to generate overly simplistic or repetitive data patterns, which could artificially inflate success rates without providing meaningful evaluation coverage. The critic component addresses this by explicitly evaluating data complexity and variety as core quality dimensions.
This improvement is reflected in the lowering of compound execution accuracy (\EXc), which quantifies how well the synthetic database can distinguish between correct and incorrect SQL queries. As shown in Figure~\ref{fig:critic-exc}, the critic's presence leads to lower \EXc~ across all three LLMs, indicating that the synthetic data is more effective at differentiating between valid and invalid queries.

This indicates that the critic component enhances not just semantic alignment with question intent,
 but also the fundamental ability to differentiate between correct and incorrect SQL queries, the core objective of robust evaluation databases.

\begin{figure}[H]
   \centering
   \includegraphics[width=1\textwidth]{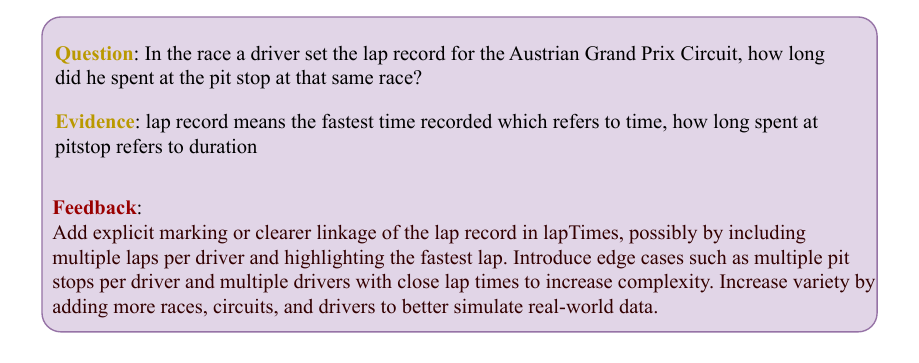} 
   \vspace{-1.5em}
      \caption{\small Example of critic feedback highlighting deficiencies in data complexity and variety, prompting regeneration of synthetic data to better align with question intent and evaluation robustness.}
      \label{fig:critic-feedback-1}
\end{figure}

\paragraph{Feedbacks}
Through detailed analysis of critic feedback across our experimental runs, we observed consistent patterns in how the critic identifies and addresses data quality issues. 
The critic provides targeted feedback such as: figure \ref{fig:critic-feedback-1} shows an example where the critic highlights deficiencies in data complexity and variety, prompting the synthesizer to regenerate synthetic data that better aligns with the question intent and enhances evaluation robustness.

\begin{figure}[H]
   \centering
   \includegraphics[width=1\textwidth]{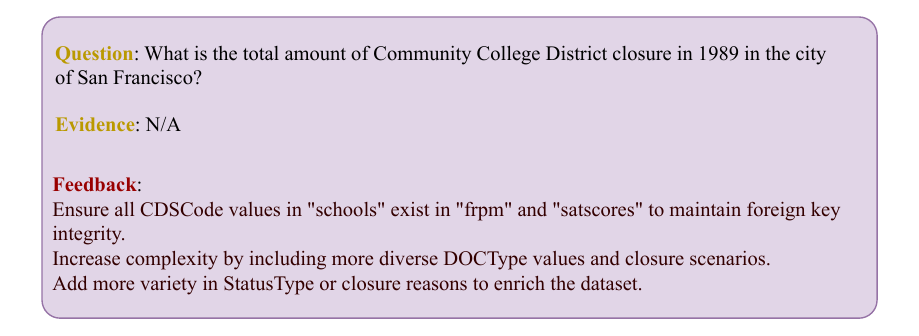} 
   \vspace{-1.5em}
      \caption{\small Example of critic feedback highlighting issues in foreign key integrity, leading to regeneration that enforces schema integrity.}
      \label{fig:critic-feedback-2}
\end{figure}

The critic also frequently identifies key integrity violations, such as non-unique primary keys or referential integrity breaches. Figure~\ref{fig:critic-feedback-2} illustrates an example where the critic detects foreign key violations, leading to regeneration that enforces these constraints and ensures schema integrity.

\begin{figure}[H]
   \centering
   \includegraphics[width=1\textwidth]{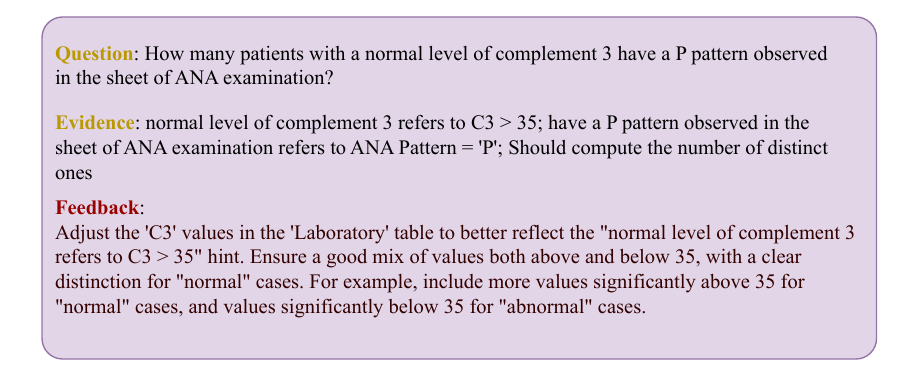} 
   \vspace{-1.5em}
      \caption{\small An example of critic feedback}
      \label{fig:critic-feedback-3}
\end{figure}

\begin{figure}[H]
   \centering
   \includegraphics[width=1\textwidth]{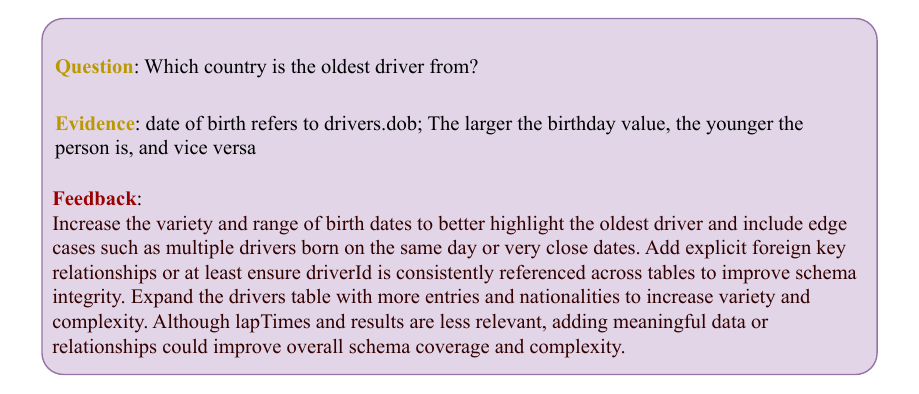} 
   \vspace{-1.5em}
      \caption{\small An example of critic feedback}
      \label{fig:critic-feedback-4}
\end{figure}

More examples of critic feedback are shown in Figures~\ref{fig:critic-feedback-3} and \ref{fig:critic-feedback-4}, demonstrating the critic's consistent role in identifying and rectifying data quality problems.

This systematic feedback mechanism ensures that subsequent iterations produce more robust test databases that can effectively distinguish between semantically correct and incorrect SQL queries.
Overall, the critic's feedback focuses on: (1) key integrity and schema coverage to ensure structural validity,
(2) presence of edge cases and boundary values, (3) diversity in categorical attributes, (4) realistic distributions that reflect real-world data patterns while aligning with question intent, and (5) inclusion of potential query failure scenarios. This multi-dimensional assessment prevents the framework from converging on overly simplistic data that might mask SQL generation errors, ensuring that high success rates reflect genuine semantic alignment and increases the robustness of evaluation, rather than artificially accommodating weak queries.

\subsection{Realism and Minimalism of Synthetic Databases}
\label{sec:appendix-realism-and-minimalism}

SynSQL-generated databases are not only realistic but also minimal and lightweight. This property is particularly valuable in scenarios where gold queries are unavailable, not only during synthesis but also for evaluation. For example, in production or cold-start settings, it is crucial to inspect and validate the generated database, either through human-in-the-loop processes or by generating expected outputs via table reasoning. The compactness of SynSQL databases facilitates such inspection and validation, making them practical for robust evaluation even when large-scale or gold-standard annotations are not accessible.

\begin{figure}[H]
   \centering
   \includegraphics[width=1\textwidth]{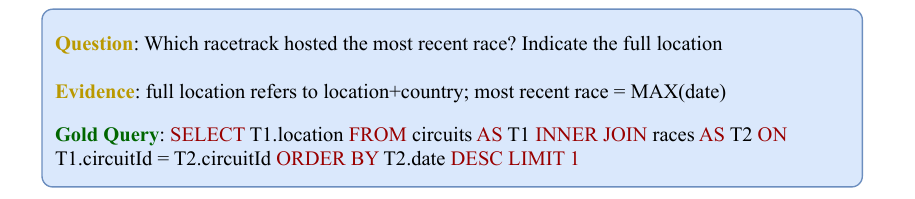} 
   \vspace{-1.5em} 
   \caption{ \small An example from the \texttt{formula\_1} database (question 1000).}
   \label{fig:db-query-1}
\end{figure}

\begin{figure}[H]
   \centering
   \includegraphics[width=0.8\textwidth]{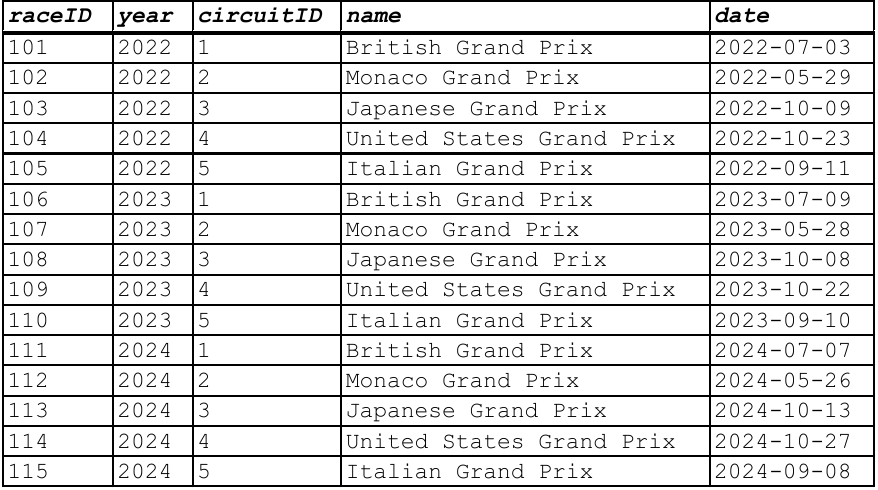} 
   \vspace{-1em} 
   \caption{\small Generated synthetic table \texttt{races} for question 1000 from the \texttt{formula\_1} database. The synthetic data contains realistic values that align with the question intent.}
   \label{fig:db-example-1}
\end{figure}

\begin{figure}[H]
   \centering
   \includegraphics[width=0.8\textwidth]{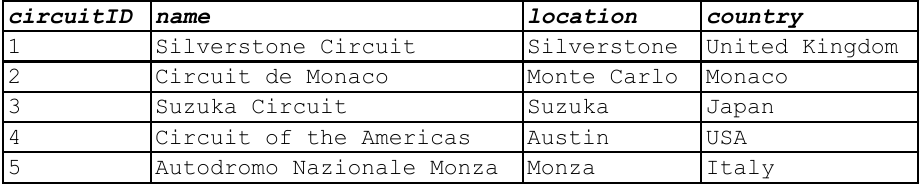} 
   \vspace{-1em} 
   \caption{\small Generated synthetic table \texttt{circuits} for question 1000 from the \texttt{formula\_1} database. The synthetic data contains realistic values that align with the question intent.}
   \label{fig:db-example-2}
\end{figure}

\vspace{1em}
\noindent
We saw in Table~\ref{tab:abstudy} that in SynSQL we have an average of 8.37 columns for BIRD and 6.71 columns for Spider to answer a query, significantly fewer than the full schemas of 75.56 and 24.55 columns respectively.
We see an example of this in question 1000 from the \texttt{formula\_1} database (see Figure~\ref{fig:db-query-1}), SynSQL generates a minimal database with only 2 tables and 9 columns, compared to the original database's 13 tables and 94 columns. The synthetic database sufficiently covers the question and relevant edge cases while being just 20KB in size, whereas the original is 21,836KB, making SynSQL's output much easier to inspect and validate. In contrast, synthetic databases generated by prior work such as TestSuiteAccuracy (TSA) often contain random values from fuzzing and are typically as large as the original databases.
As illustrated in Figures~\ref{fig:db-example-1} and~\ref{fig:db-example-2}, which show the entirety of data generated for this question by SynSQL, the synthetic data includes realistic values that closely match the question intent.

Moreover, SynSQL ensures that values within each row are meaningfully related and contextually accurate. For example, if a row in the \texttt{races} table has the year set to \texttt{2024}, all corresponding data in that row (such as race name or date) is consistent with that year. Similarly, in the \texttt{circuits} table, if the location is \texttt{Monza}, the country is set to \texttt{Italy}, reflecting the real-world fact that there is a Formula 1 Grand Prix held in Monza, Italy. This level of realism and consistency, both within rows and across related tables, is achieved by leveraging LLMs to generate data that maintains semantic coherence and factual alignment.

\label{sec:appendix-inconsistencies-examples}
\subsection{Inconsistency Examples from Spider Dev Set}

\begin{figure}[h]
   \centering
   \includegraphics[width=1\textwidth]{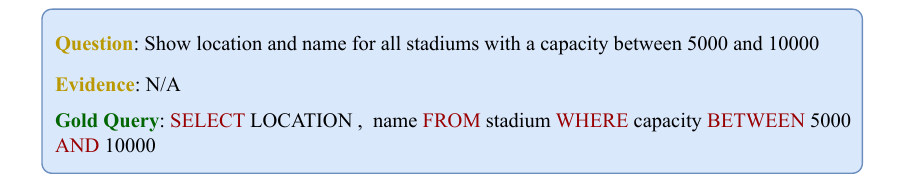} 
   \vspace{-1.5em} 
   \caption{\small An example of inconsistencies between gold query and database contents in the Spider dev set. SynSQL aligns with the question, leading to recovery of such inconsistencies.}
   \label{fig:sr-example-6}
\end{figure}

\begin{figure}[h]
   \centering
   \includegraphics[width=1\textwidth]{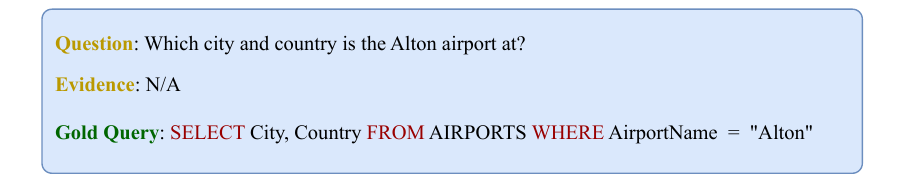} 
   \vspace{-1.5em} 
   \caption{\small An example of inconsistencies between gold query and database contents in the Spider dev set. SynSQL aligns with the question, leading to recovery of such inconsistencies.}
   \label{fig:sr-example-7}
\end{figure}

There are questions in the spider dev set that the gold query does not align with the content of original test databases. Below are some examples of such inconsistencies, which lead to the observed low success rates for the original Spider databases. SynSQL generates synthetic data that aligns with the question intent and recovers such inconsistencies.
For example, in (Figure~\ref{fig:sr-example-6}) the questions asks for the location and name for all stadiums with a capacity between 5000 and 10000. However, there are no such stadiums in the original database, leading to the gold query returning empty results. SynSQL generates synthetic data that includes stadiums within this capacity range.
Another example is shown in (Figure~\ref{fig:sr-example-7}), where the question asks for the city and country of the Alton airport. However, the original database \texttt{flight\_2} has the airport name listed as \texttt{Alton  },  with a trailing space, leading to a mismatch with the gold query. SynSQL generates synthetic data that correctly matches the airport name as specified in the question.

\subsection{Column Selection Prompt}
\label{sec:prompt-colSel}
\begin{figure}[H]
   \centering
   \includegraphics[width=0.9\textwidth]{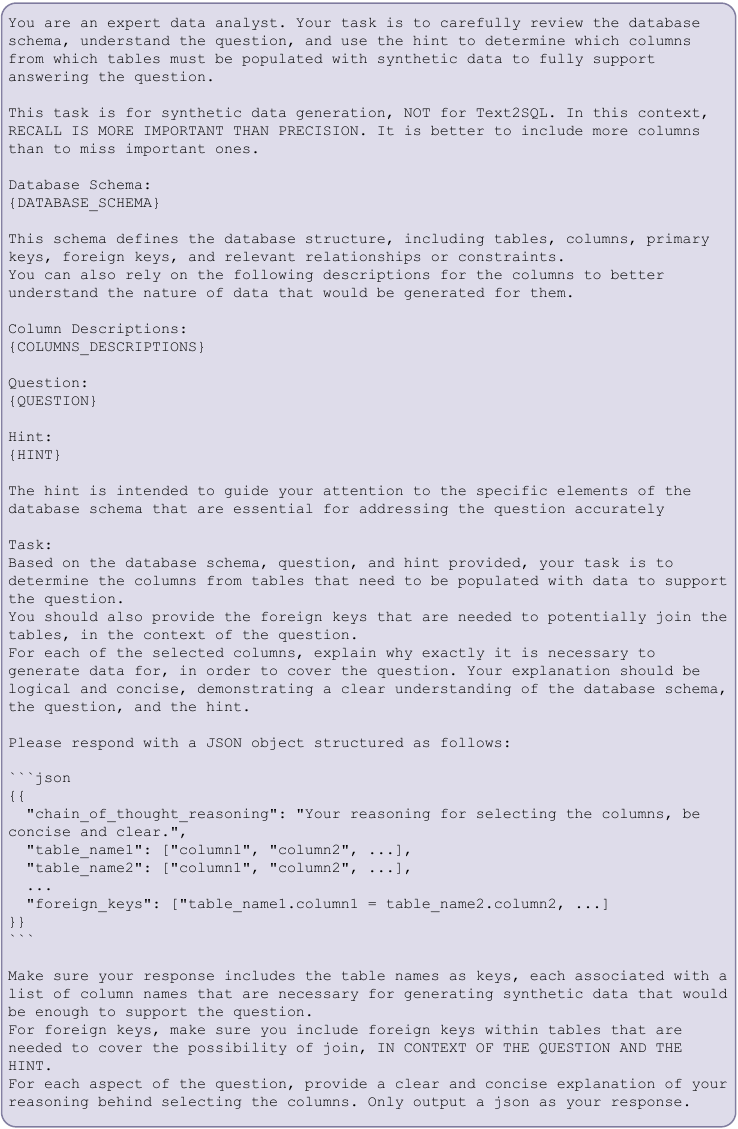} 
   \caption{\small The prompt template used for column selection in the schema selector component of SynSQL.}
   \label{fig:prompt_1}
\end{figure}

\newpage

\subsection{Column Expansion Prompt}
\label{sec:prompt-colExp}
\begin{figure}[H]
   \centering
   \includegraphics[width=0.9\textwidth]{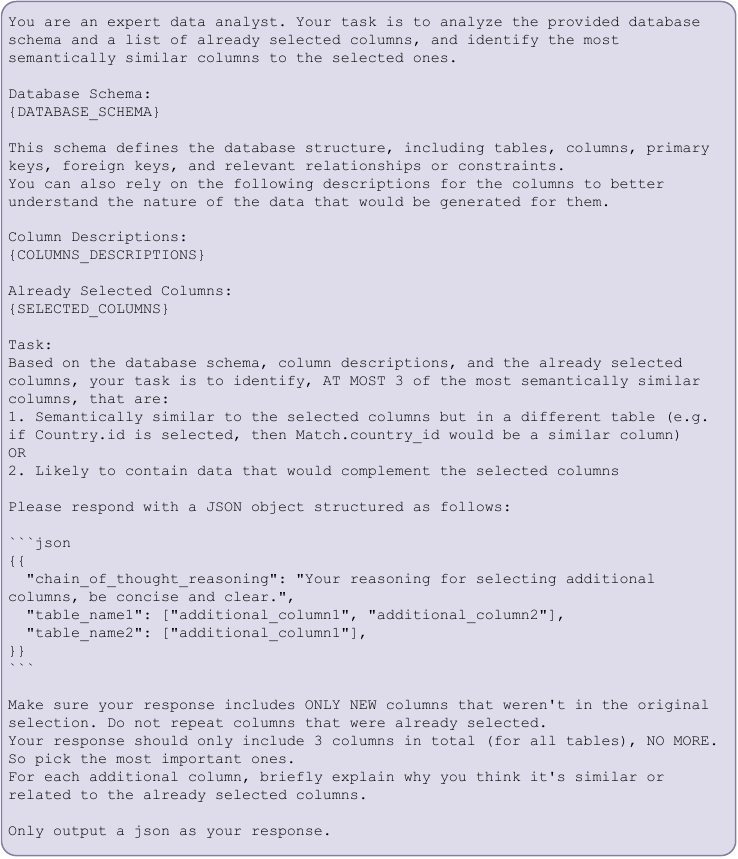} 
   \caption{\small The prompt template used for column expansion in the schema selector component of SynSQL.}
   \label{fig:prompt_2}
\end{figure}

\newpage

\subsection{Data Synthesizer Prompt}
\label{sec:prompt-synthesizer}
\begin{figure}[H]
   \centering
   \includegraphics[width=0.9\textwidth]{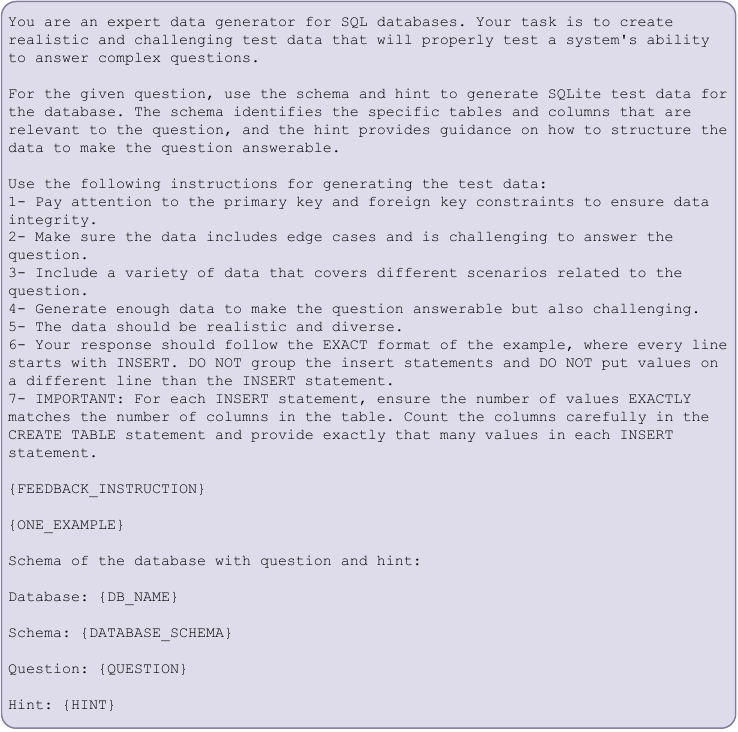} 
   \caption{\small The prompt template used for data synthesis component of SynSQL.}
   \label{fig:prompt_3}
\end{figure}

\newpage

\subsection{Data Critic Prompt}
\label{sec:prompt-critic}
\begin{figure}[H]
   \centering
   \includegraphics[width=0.9\textwidth]{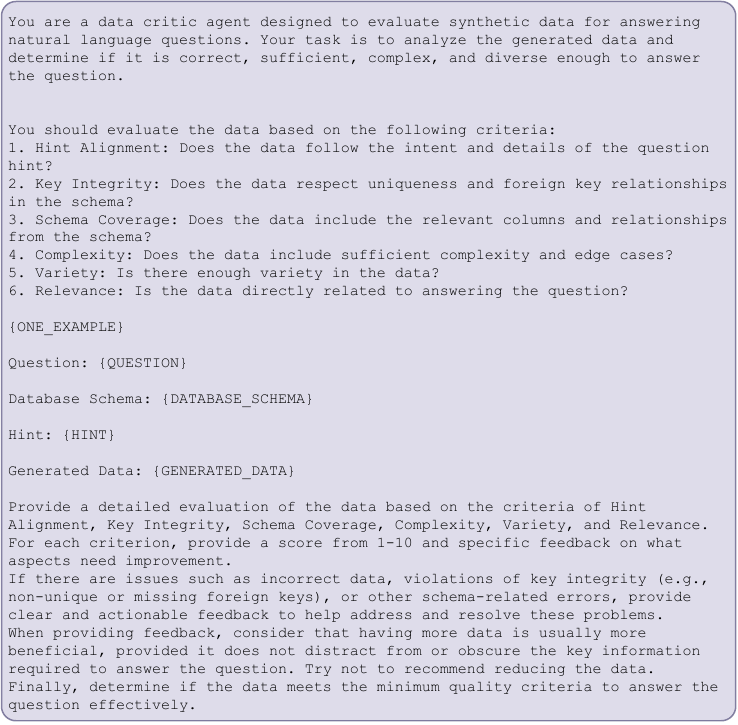} 
   \caption{\small The prompt template used for data critic component of SynSQL.}
   \label{fig:prompt_4}
\end{figure}


\end{document}

%% file: math_commands.tex
\usepackage{amsmath,amsfonts,bm}

\def\eqref#1{equation~\ref{#1}}

\def\1{\bm{1}}

\DeclareMathAlphabet{\mathsfit}{\encodingdefault}{\sfdefault}{m}{sl}
\SetMathAlphabet{\mathsfit}{bold}{\encodingdefault}{\sfdefault}{bx}{n}

